\documentclass[conference]{IEEEtran}
\IEEEoverridecommandlockouts
\usepackage{cite}
\usepackage{amsmath,amssymb,amsfonts}
\usepackage{algorithmic}
\usepackage{graphicx}
\usepackage{subcaption}
\usepackage{textcomp}
\usepackage{xcolor}
\usepackage{amsmath}
\def\BibTeX{{\rm B\kern-.05em{\sc i\kern-.025em b}\kern-.08em
    T\kern-.1667em\lower.7ex\hbox{E}\kern-.125emX}}
\begin{document}

\title{A Novel PHY Layer Approach for Enhanced Data Rate in LoRa using Adaptive Symbol Periods  \\
}

\author{\IEEEauthorblockN{Akshay Ramesh Jadhav}
\IEEEauthorblockA{\textit{Department of Electrical Engineering} \\
\textit{Indian Institute of Technology Hyderabad}\\
Hyderabad, INDIA \\
ee16resch01002@iith.ac.in}
\and
\IEEEauthorblockN{P. Rajalakshmi}
\IEEEauthorblockA{\textit{Department of Electrical Engineering} \\
\textit{Indian Institute of Technology Hyderabad}\\
Hyderabad, INDIA \\
raji@iith.ac.in}
}

\maketitle

\begin{abstract}
LoRaWAN has emerged as one of the promising low-power wide-area network technologies to enable long-range sensing and monitoring applications in Internet of Things. 
The LoRa physical layer used in LoRaWAN suffers from low data rates and thus increases packet duration. 
%Due to a large number of devices deployed to form a dense network and their use of simple media access protocols, the collision performance and 
%network throughput degrades as packet duration increases due to lower data rates.
In a dense LoRaWAN network scenario with simple media access protocol like ALOHA, the packet collision probability increases with increase in packet duration.
This degrades over-all network throughput because of increased re-transmissions of collided packets. 
Any increase in data rate directly reduces the packet duration.
Thus, in this paper, we have proposed a novel approach to enhance the data rate in LoRa communication system by using adaptive symbol periods in physical layer. 
To the best of our knowledge, this is the first attempt at using adaptive symbol periods to enhance data rate of the LoRa system. 
The trade-off of the proposed approach in terms of required symbol overhead and degradation in bit error rate performance due to symbol period reduction has also been analysed. 
% We show that for reduction factor \(\beta = 0.875\), the data rate directly increases \(1/\beta = 1.1428\) times.
We have shown that for reduction factor \(\beta\), the data rate directly increases \(1/\beta\) times.
% , while degrading signal-to-noise ratio by less than \(1 dB\). 
%Its impact is shown to be negligible for specific classes of devices in LoRa.

\end{abstract}

\begin{IEEEkeywords}
Internet of Things, Low-Power Wide-Area Network, LoRa, Chirp Spread Spectrum, Physical Layer
\end{IEEEkeywords}

\section{Introduction}

Low-power wide-area network (LPWAN) protocols play an important role in Internet of Things (IoT) by providing long-range communication while keeping transmission energy low \cite{LPWAN} \cite{LPWAN1}. 
This translates to increased lifetime of several years for battery operated IoT devices.
LPWAN also suffers from low data rates below just a few kbps. 
This small data rate is suitable for applications specific to asset tracking, structural health monitoring, energy metering in smart grid, smart city infrastructure, etc.

LPWAN enables a large number of IoT devices to be deployed in a small area, making up a very dense IoT network. 
These devices are supposed to be constrained in hardware capabilities as well as energy budget for data transmission. 
Thus, here arises requirement for a highly sophisticated and advanced media access control (MAC) protocol to accommodate large number of devices in a dense network. 
However, due to processing and energy constraints, MAC protocol being used in such networks is often as simple as ALOHA or time division multiple access (TDMA). 
%Distributed Queuing approach instead of aloha

LoRaWAN is one of the LPWAN technologies, which has gathered much attraction in recent years. 
Its specifications are maintained by LoRa Alliance \cite{lorawan}\cite{alliance}. 
LoRa is the physical layer modulation scheme used in the LORAWAN. 
It has been patented by Semtech Corporation \cite{p1}.
Other LPWAN technologies which are also actively being used are NBIoT, SigFox, etc. \cite{ded}.
In this paper, we focus on LoRaWAN technology to improve its data rate by adaptive symbol period reduction of LoRa modulation scheme.
This further improves the collision performance and thus improves overall network throughput. 

Chirp spread spectrum (CSS) \cite{chirp} is used in the LoRa physical layer because of its resilience to noise and good sensitivity \cite{fscm}. 
The main principle of CSS is to use of much larger transmission bandwidth than required.
% , which grants it these useful properties.
Due to this, the receiver can achieve higher sensitivity and recover signals over a longer distance even at poor SNR conditions.
% This comes however at the cost of lower data rate.

%Lora physical layer uses frequency shifted chirp spread spectrum. All possible cyclic time shifts of the base upchirp correspond to all possible symbols. At the receiver side, the rf sampled data is mutiplied with conjugate of the base upchirp i.e. downchirp. This process transforms the signal to a sinusoidal signal with frequency dependant on amount of cyclic shift in modulated upchirp.

%In rf channel, multiple spreading factors (SF) can be used. These SFs are shown to be quasi-orthogonal in the literature.
%Different SFs provide different data rats and ranges.
%Due to limited frequency bands and limited number of SFs defined by LORAWN alliance, there comes a limit to total network throughput.
%This maximum throughput however can not be achieved due to loss of 
%In addition to this, the slotted aloha mechanism used by LORAWAN

However, LoRa suffers from very low data rates, thus increasing the packet duration.
% Packet duration is number of symbols multiplied by time period of one symbol.
Due to use of ALOHA based MAC protocol, the higher packet duration translates to increase in the packet collision probability and thus degradation of network throughput \cite{collide}. 

%By decreasing packet duration, collision probability can be decreased and network throughput can be improved. 

In this paper, we have proposed a novel approach for data rate enhancement using adaptive symbol periods. 
Increased data rate decreases the packet duration.
This helps in achieving better collision performance as the individual packet duration decreases. 
Since the transmission energy is also directly proportional to packet duration, reduction in packet duration can also save transmission energy of the end device.
To the best of our knowledge, this kind of approach has never been explored in the existing literature. 
Therefore this is the first time adaptive symbol period in LoRa for data rate enhancement has been proposed.

Rest of the paper is organized as follows. In Section II, we provide a brief introduction to a few basics of LoRaWAN which are important to understand our proposed approach.
LoRa physical layer is explained in detail in Section III. 
The proposed approach of data rate enhancement in LoRa using adaptive symbol period is explained in Section IV. 
Results and discussion regarding merits and trade-offs of the proposed approach are given in Section V. 
Conclusion of the paper is given in Section VI.

%\section{Related Work}
%
%In the existing literature, many improvements are continuously being proposed for the LoRaWAN system. Most of the proposed approaches are trying to optimize the parameters defined by LoRaWAN specification.
%Adaptive data rate mechanism and its improvements are explained in \cite{config}.
%Researchers are already trying to improve lora efficiency in order to accommodate more devices and also to increase overall network throughput (Date Extraction Rate).
%Some have proved that putting extra base stations and putting directional antennas on end devices can be helpful in increasing network capacity.
%
%In current scenario of LoRaWAN deployments and specifications, if two transmissions are collided, then the end devices can identify collision through ACK-NACK procedure and re-transmit the packet again. This adds up to the transmission energy consumption of end device as well as reduction in over-all network data throughput.
%
%Lora physical layer also has some chance of collision recovery through capture effect. Because of this effect, out of two overlapping signals on same channel and same SF, one can be successfully decoded if it is sufficiently high in power compared to the other.

\section{Primer To LoRaWAN}

LoRaWAN network consists of end devices communicating with a central gateway operating in a star topology \cite{lorawan}. Based on the device transmission and reception capabilities, LoRaWAN specification \cite{alliance} has defined three categories of devices. 
\begin{itemize}
\item Class A: After each uplink transmission, devices of this category can receive downlink messages in two short reception windows. 
Other than these times, the device stays in sleep mode to reduce the energy consumption. 
Thus, these devices can be operated with a battery and can last for several years.
% Thus these devices waste minimum energy for reception. 
\item Class B: Devices in this category have a predetermined periodic reception window. 
These devices can receive periodic beacons from gateway server and thus, can be controlled from gateway with latency equal to the beacon interval. 
These devices have higher energy consumption than Class A devices.
\item Class C: Devices of this category have their reception always on, except for the time of transmission. 
% These devices consume the most energy compared to Class A and Class B devices. 
These devices do not have any sleep window and thus, their energy consumption can not be optimized.
However, because of always active reception, these devices offer the minimum downlink latency.
\end{itemize}

Since an IoT device needs to have lower energy consumption and increased battery lifetime, Class A devices are more popular and more widely used than other classes. 
These devices provide battery lifetime of several years along with a very wide communication range. This makes these devices very much suitable for IoT applications.

%However, LoRaWAN network can also have Class B and Class C devices, which are not as energy constrained as Class A. Therefore, for Class B and Class C devices, improving data rates is more important than optimizing transmission energy. This will be further discussed in section V.
However, LoRaWAN network can also have class C devices, which are not as energy constrained as class A and class B devices. 
For such devices, improving data rate can be more beneficial towards mitigating the network congestion, than optimizing transmission power. 
This is further discussed in section V.

\subsection{LoRa Packet Structure}
LoRa physical layer packet consists of preamble, header and payload symbols.
The preamble serves the purpose of timing synchronization for correct recovery of the symbols. 
It consists of a sequence of unmodulated upchirp symbols, followed by 2.25 symbols of unmodulated downchirp for notifying the receiver of start of header and payload.

After preamble, header symbols are transmitted which contain information about length of packet, coding rate and presence of CRC of payload along with its own CRC.
In Implicit Header mode, the header can be removed to reduce packet length and decrease packet duration. 
In this case, both transmitter and receiver need to be in agreement of these parameters.
% In order to provide minimum probability of error, the header is always transmitter at 4/8 coding rate which gives maximum error correction capability at the receiver.
Header is followed by payload symbols and payload CRC. 
% The maximum number of symbols in payload is limited to 255.

\subsection{LoRaWAN Transmission Options}
According to LoRaWAN specification \cite{alliance}, there are five configuration parameters which can be adjusted to optimize the network. 
These parameters are as following,

\begin{itemize}
\item Carrier Frequency (CF):
CF is the center frequency of the frequency variation in the chirp signal. According to specification, CF can be chosen from 137 MHz to 1020 MHz. 
Since some parts of these frequencies may be unavailable in different regions of the world due to licensing, it is advised to use frequencies according to the region of operation.

\item Bandwidth (BW):
Bandwidth is the difference between minimum and maximum frequencies of the chirp signal. 
According to specifications, BW of 125 kHz, 250 kHz or 500 kHz can be used. 
Higher BW gives higher data rate, but at the same time it also reduces the receiver sensitivity.

\item Spreading Factor (SF):
According to specifications, SF can be chosen from 7 to 12.
SF determines the number of chips per symbol which is given as \(2^{SF}\).
Higher SF increases the sensitivity of receiver against noise and thus range of communication. 
But increasing SF by one also doubles the time per symbol, and thus halves the symbol transmission rate.

\item Transmission Power:
The transmission power of the LoRa radio module can be varied in steps from -4 dB to 20 dBm. 
The adaptive variation of transmission power helps in reducing transmission energy at battery operated end device and thus extending its lifetime.

\item Coding Rate (CR): 
To improve bit error rate (BER) performance in presence of burst interference, a forward error correction (FEC) scheme is used. 
According to LoRaWAN specifications, CR of 4/5, 4/6, 4/7 or 4/8 can be used. 
CR of 4/8 offers highest redundancy and thus highest resilience to noise, but also doubles the time-on-air compared to the un-coded packet.

\end{itemize}

\subsection{Adaptive Data Rate (ADR) Mechanism}
ADR mechanism is an important part of LoRaWAN specification \cite{config}. 
It enables the system to optimize spreading factor (SF) and transmission power (TP) of all end devices in order to increase reliability of communication as well as optimize energy consumption of battery powered end devices\cite{adr}.

If an end device does not receive any acknowledgement on downlink after multiple consecutive uplink transmissions, then it assumes that its transmissions are not being decoded properly at the receiver, i.e. the connection has been lost. 
The end device tries to adapt to this change by gradually increasing its TP to the maximum possible. 
After maximum possible TP has been achieved and still no acknowledgment is being received, the end device increments its SF by 1.
The end devices thus change their SF \& TP values to establish a reliable uplink. 
This may not be an efficient communication link as energy consumption of end device increases as TP increases \cite{agile}.

Thus, there is also provision for the LoRa network server to monitor uplink quality and command end devices to change their SF \& TP values in order to maximize the network throughput.

The uplink quality of each packet of each end device is stored in history of network server. 
If the uplink quality for recent packets of an end device is found to be much better than the minimum required, the network server can notify that end device to reduce its transmission power and SF value.
The purpose of this adjustment is to ensure that the signal-to-noise ratio (SNR) of uplink is not too much greater than the minimum required.

The reduction in SF increases data rate and thus decreases the transmission time of packet (time-on-air). 
Reduction in TP reduces transmission energy and thus increases lifetime of the battery operated end device.

\section{Lora Physical Layer}

Lora physical layer consists modulation of symbols, transmission of modulated signal and demodulation of symbols by the receiver.
LoRa physical layer uses CSS based modulation technique \cite{semtech}. 
A chirp is a sinusoidal signal whose frequency is linearly varying over time. 
Thus it is also called as linear frequency modulated (LFM) signal. 
Since the signal is spread over a wide bandwidth, the sensitivity at the receiver increases and the signal with poor SNR can also be recovered successfully.

\subsection{LoRa Modulation Scheme}

Assume the minimum and maximum instantaneous frequencies of the chirp to be \(0\) and \(F\) Hertz. Thus bandwidth of chirp becomes \(F\) Hertz. 
Thus for time per symbol \(T_s\), the rate of increase of frequency of an upchirp can be written as \(a = F / T_s\).
For downchirp, the rate of decrease of frequency can be written as (\(- a\)) \( = - F / T_s\).

\begin{figure}[b]
\begin{center}
\centering
\begin{subfigure}{.23\textwidth}
\includegraphics[width=1.79in]{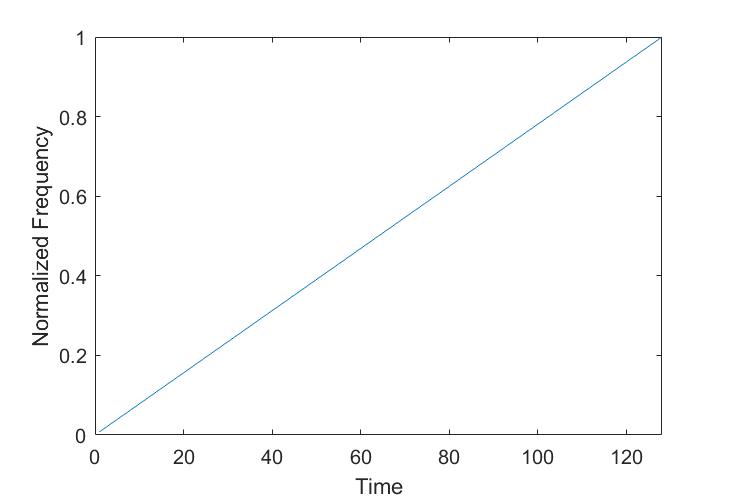}
\caption{{Upchirp}}
\label{fig:u}
\end{subfigure}
\begin{subfigure}{.23\textwidth}
\includegraphics[width=1.79in]{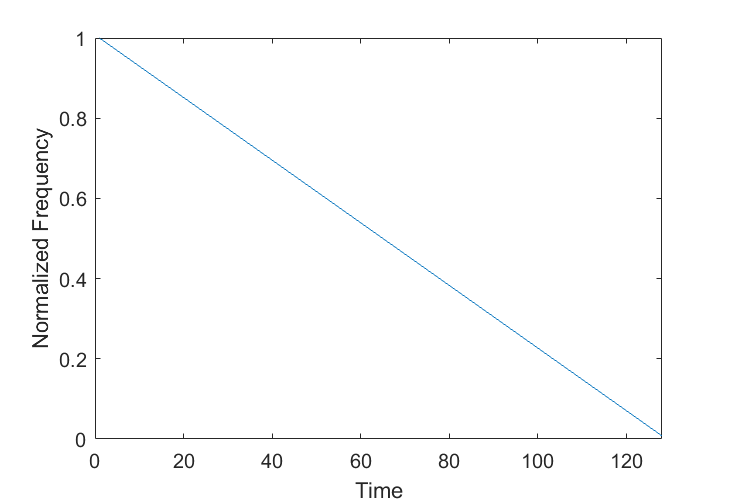}
\caption{\textsf{Downchirp}}
\label{fig:d}
\end{subfigure}
\caption{{Frequency variation for one symbol period of upchirp \& downchirp}}
\end{center}
\end{figure}

The basechirp used for modulation of payload symbols is an upchirp, i.e. a sinusoidal signal whose frequency is linearly increasing from \(0\) to \(F\) Hz as shown in Fig.~\ref{fig:u}. 
Thus the base upchirp \(U_0(t)\) and its frequency variation \(F_{U0}(t)\) can be written as following.

\begin{equation} \label{u0}
  U_0(t)=e^{j\pi(at)t} \quad \text{for } 0\leq t\leq {T_s}.
\end{equation}

\begin{equation} \label{fu0}
  F_{U0}(t)={(at)} \quad \text{for } 0\leq t\leq {T_s}.
\end{equation}

Similarly, a downchirp is a sinusoidal signal whose frequency is linearly decreasing from \(F\) to \(0\) Hz as shown in Fig.~\ref{fig:d}. 
Thus, the downchirp \(D_0(t)\) and its frequency variation \(F_{D0}(t)\) can be written as following.

\begin{equation} \label{d0}
D_0(t) = e^{j\pi(F - at)t} \quad \text{for } 0\leq t\leq {T_s}.
\end{equation}

\begin{equation} \label{fd0}
  F_{D0}(t)={(F - at)} \quad \text{for } 0\leq t\leq {T_s}.
\end{equation}

For modulation, each LoRa symbol is encoded as a cyclic shift in the base upchirp. 
It can also be viewed as frequency shift chirp modulation \cite{fscm} as it works on the principle of modulating data symbol in terms of cyclic shifts of the upchirp. 
Spreading Factor \(SF\) determines the number of all possible symbols which is given as \(2^{SF}\). 
In other words, for a LoRa symbol \(k\), this can be written as \(k \in \{0,\dotsc,2^{SF}\}\).
Thus, in the total time period for one symbol \(T_s\), one of \(2^{SF}\) possible cyclic shifts needs to be used depending on the LoRa symbol being modulated.
The cyclic shifts are implemented in steps of \(T_{chip}= 1 / BW\).
%The difference between consecutive cyclic time shifts is \(T_{chip}= 1 / BW\). 

\begin{figure}[h]
\begin{center}
\centering
\includegraphics[width=2.5in]{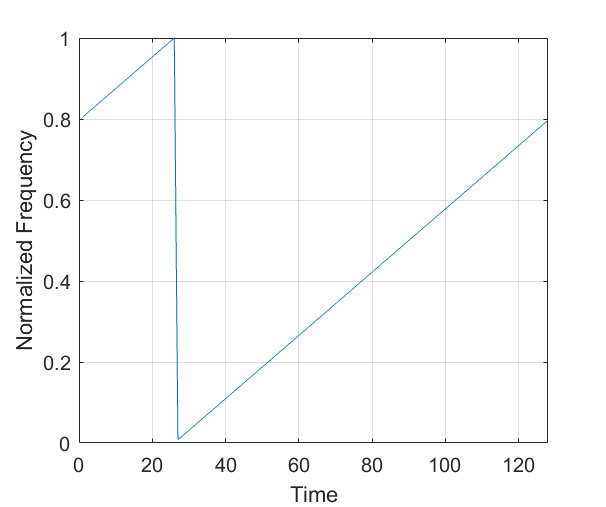}
\caption{{Frequency variation for one symbol period of cyclic shifted upchirp}}
\label{fig:uk}
\end{center}
\end{figure}

Let \(T_k = k \times T_{chip}\) be the cyclic shift corresponding to symbol \(k\).
Therefore, corresponding to symbol \(k\), the cyclic shifted upchirp \(U_k(t)\) and its frequency variation \(F_{U0}(t)\) (refer Fig.\ref{fig:uk}) can be written as following.

\begin{equation}\label{uk}
U_k(t) =
\begin{cases} 
      e^{j\pi(F + a(t-T_k))t} & \text{for }0\leq t\leq T_k, \\
      e^{j\pi(a(t-T_k))t} & \text{for }T_k\leq t\leq T_s.
   \end{cases}
\end{equation}

\begin{equation}\label{fuk}
F_{Uk}(t) =
\begin{cases} 
      {(F + a(t-T_k))} & \text{for }0\leq t\leq T_k, \\
      {(a(t-T_k))} & \text{for }T_k\leq t\leq T_s.
   \end{cases}
\end{equation}

The modulated upchirp corresponding to each symbol is transmitted for time period \(T_s\).
Thus the total transmission time for \(N_s\) symbols, i.e. time-on-air becomes \(N_s \times T_s\).

\begin{figure*}[t]
\begin{center}
        \centering
        \includegraphics[width=7.5in]{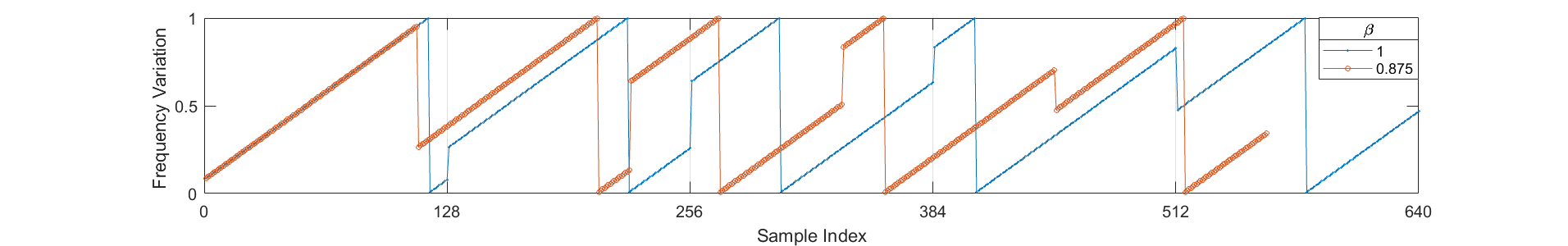}
        \caption{{Normalised frequency variation of five LoRa symbols for reduction factor \(\beta=1\) and \(0.875\) (symbol rate increased to \(1.143\) times).}}
        \label{fig:beta}
\end{center}
\end{figure*}

\subsection{LoRa Demodulation Scheme}
In literature as well as in practice, the orthogonality of LoRa chirp signals has been well established and tested.
Thus at the receiver, the best way to detect the modulated symbol i.e. the cyclic shift in the modulated upchirp is using an optimum filter matched to the frequency spectrogram of base upchirp. 

In frequency domain, upchirp and downchirp are time mirrored forms of each other. Therefore, correlation with an upchirp is same as convolution with a downchirp.
Since convolution in frequency domain is same as multiplication in time domain, the whole process of symbol detection simplifies to a simple multiplication of incoming signal with a downchirp.
In literature, this is also referred as De-Chirping of signal. 
The product of modulated upchirp \(U_k(t)\) and downchirp \(D_0(t)\) can be written as

\begin{equation}
U_k(t) \times D_0(t) = \\
\begin{cases} 
      e^{j\pi(2F -aT_k )t} & \text{for }0\leq t\leq T_k, \\
      e^{j\pi(F-aT_k)t}  & \text{for }T_k\leq t\leq T_s.
\end{cases}
\end{equation}

% \begin{equation}
% \begin{align*}
% U_k(t) &\times D_0(t) = \\ 
% &
% \begin{cases} 
%       e^{j2\pi(F + a(t-T_k))t} \times e^{j2\pi(F - at)t} & \text{for }0\leq t\leq T_k \\
%       e^{j2\pi(a(t-T_k))t} \times e^{j2\pi(F - at)t} & \text{for }T_k\leq t\leq T_s 
%   \end{cases}\\
% =&
% \begin{cases} 
%       e^{j2\pi(2F -aT_k )t} & \text{for }0\leq t\leq T_k \\
%       e^{j2\pi(F-aT_k)t}  & \text{for }T_k\leq t\leq T_s
%   \end{cases}
% \end{align*}
% \end{equation}

When sampled at sampling frequency of \(F\) Hz, the terms \(2F\) and \(F\) in the above equation vanish and therefore can be neglected in the analysis.

After the received signal is de-chirped by multiplying with a downchirp, the product gives a complex sinusoidal signal with frequency  proportional to the cyclic time shift \(T_k\) in the modulated upchirp.

This frequency can be calculated by taking \(2^{SF} \) point Inverse Fast Fourier Transform (IFFT) of the product. 
The symbol can be found by searching for the index of peak in the IFFT magnitude.

The margin by which the magnitude of this maximum peak in the IFFT rises above the noise-floor is also a measure of SNR. 
If the SNR is good, the peak is higher with respect to noise-floor and can be easily detected by the receiver. 
Similarly, if the SNR is poor, the peak is just above the noise-floor, making it difficult to be detected.

\section{The Proposed Approach for Adaptive Symbol Period in LoRa PHY Layer}

Our intuition to reduce the symbol period comes from the fact that when SNR is high, some samples in the received modulated upchirp can be strategically dropped without loosing ability to detect the required peak in IFFT.

In the proposed approach, for each symbol of payload, instead of transmitting the corresponding cyclic shifted upchirp for the total symbol period \(T_s\), it is transmitted only for period \( T'_s = \beta \times T_s\), where \(\beta \in (0,1]\)  is the reduction factor.

\begin{equation}\label{uk1}
U'_k(t) =  U_k(t) \quad \text{for }0\leq t\leq T'_s. 
\end{equation}

The transmitter then starts to transmit next symbol's upchirp after time period \( T'_s\).
Thus the modulated upchirp corresponding to each symbol is transmitted for time duration  \(T'_s\) only, where \(T'_s \leq T_s\). 

As seen in Fig.~\ref{fig:beta}, \(\beta = 1\) corresponds to symbol period \(T_s = 128\) samples and \(\beta = 0.875\) corresponds to reduced symbol period \( T'_s = 0.875 \times 128 = 112\) samples. 
Thus, for the same 5 symbols, the proposed approach requires 560 samples compared to 640 original samples.
The proposed approach with \(\beta = 1\) improves symbol rate \(1 / 0.875 = 1.143\) times the original symbol rate.

\begin{figure}[t]
\begin{center}
\centering
\includegraphics[width=3.5in]{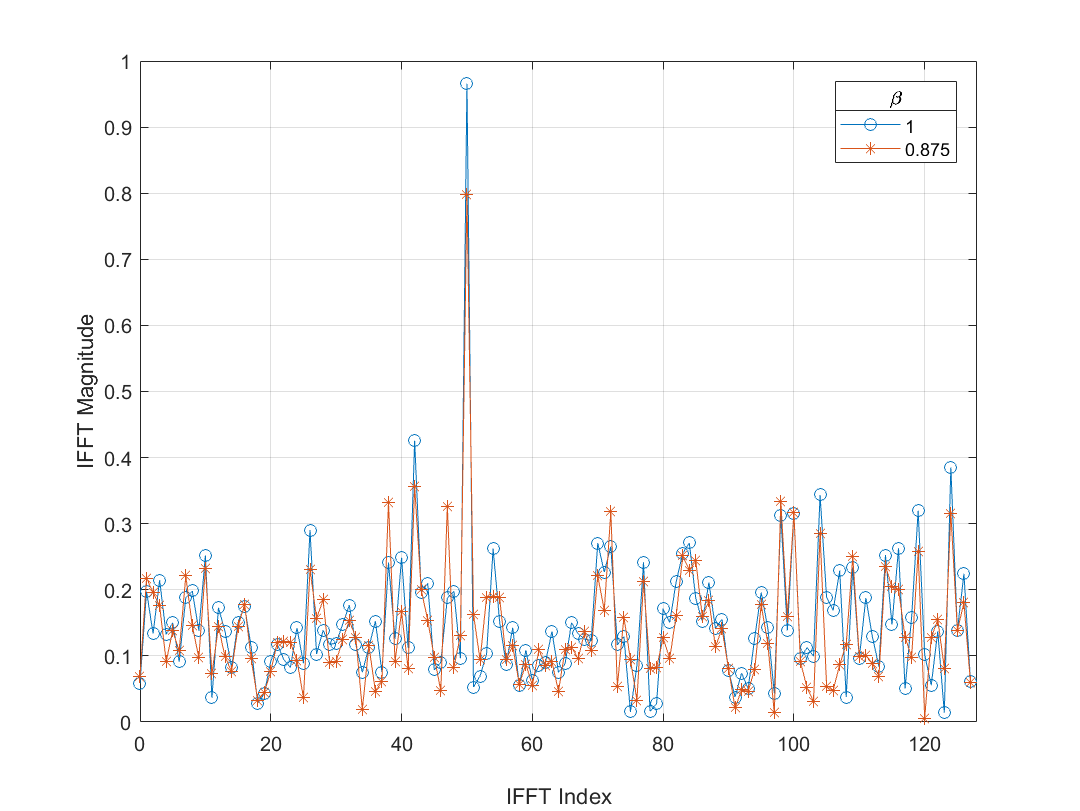}
\caption{{IFFT Magnitude for \(\beta=1\) and \(\beta=0.875\); Signal power is reduced to 80\% of the original. }}
\label{fig:875}
\end{center}
\end{figure}

In Fig.~\ref{fig:875} \& Fig.~\ref{fig:50}, IFFT magnitude of de-chirped signal for a single LoRa symbol is plotted for \(\beta\) values of \(0.875\) \& \(0.5\). 
The simulations are performed assuming an additive white Gaussian noise of \(0 dB\).
It can be seen that the lower reduction factor also decreases symbol energy, thus degrading signal-to-noise ratio. 
This decreases the detection probability of correct symbol for lower values of reduction factor.

Our solution is to use adaptive reduction factor \(\beta\) depending on the SNR conditions of the communication link. 
The receiver needs to know value of the adaptive reduction factor \(\beta\) to determine \(T'_s\) in order to correctly decode the symbols.
We address this issue by introducing an extra symbol in the header part of the packet. This symbol is to convey information about \(\beta\) value used by transmitter for subsequent payload symbols.

Since preamble and header parts are critical for packet detection as well as frequency and time synchronization, their symbol periods should not be modified. 
This ensures that the preamble detection at the receiver is not degraded due to the proposed approach.

For decoding procedure for the proposed approach, receiver can use information about \(\beta\) from header to calculate \(T'_s\). Accordingly, after end of header, consecutive received signal blocks of period \(T'_s\) can be used for decoding corresponding symbols of payload. 

Therefore, at the receiver, \(U'_k(t)\) is multiplied by truncated downchirp i.e. \(D'_0(t)\). 

\begin{equation}\label{uk1}
D'_0(t) =
\begin{cases} 
      D_0(t) & \text{for }0\leq t\leq T'_s, \\
      0 & \text{for }T'_s\leq t\leq T_s.
   \end{cases}
\end{equation}

The element-by-element product of \(U'_k(t)\) and \(D'_0(t)\) is equivalent to \(U_k(t) \times D_0(t)\) truncated at \(T'_s\).

\begin{equation}\label{uk1}
U'_k(t) \times D'_0(t) = U_k(t) \times D_0(t) \quad \text{for }0\leq t\leq T'_s .
\end{equation}   

This truncation of samples in the product does not affect its frequency. 
Thus after IFFT, we can detect the correct maximum peak corresponding to symbol. 
The only change is decrease in magnitude of this peak, resulting in degradation of SNR. This effect on SNR can be observed in Fig.~\ref{fig:875} \& Fig.~\ref{fig:50}.

\section{Results and Discussion}

\subsection{Advantages of The Proposed Approach}

The proposed approach has a cascading effect on network performance. It affects multiple performance parameters of the network like energy efficiency, data rate, congestion, packet delivery rate, etc. This effect can be explained step-by-step as follows.

\begin{itemize}
\item The proposed approach decreases the symbol period by factor \(\beta\). This reduction in symbol period directly increases the physical layer data rate (\(1 / \beta\)) times.

\item Increase in data rate effectively decreases transmission time, i.e. time-on-air, of the packet.
\item Decrease in time-on-air also means a proportional decrease in transmission energy. 
This helps in increasing the lifetime of battery operated devices.
% \item Since for the same amount of information in packet, packet duration is being reduced, this will help in reducing the congestion of the network.
\item As packet duration and packet collisions are closely related, the proposed approach also reduces packet collision probability and thus increases packet delivery rate. 
Therefore it also improves the overall network throughput.
%\item As another use case of proposed approach, keeping the packet time same as original, end devices can transmit more information due to increased symbol rate. This will increase overall network throughput, but it will not give any of the other improvements like energy saving and reduction of collision probability.
\end{itemize}

\begin{figure}[t]
\begin{center}
\centering
\includegraphics[width=3.45in]{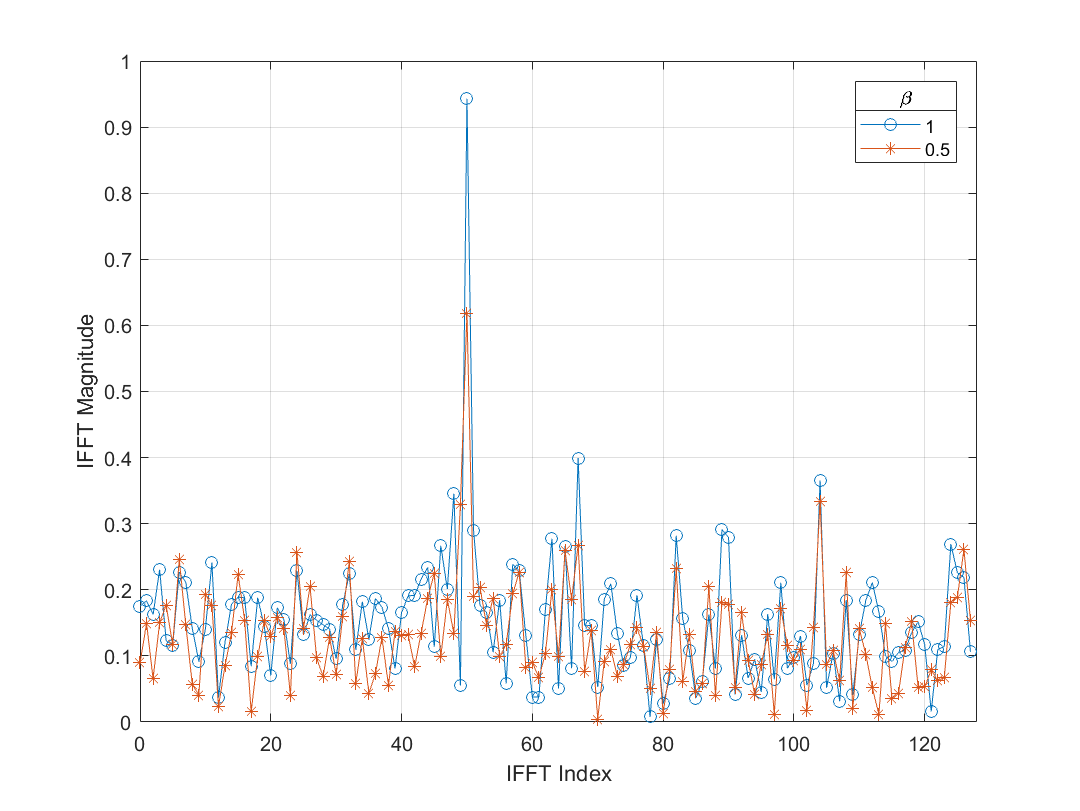}
\caption{{IFFT Magnitude for \(\beta=1\) and \(\beta=0.5\); Signal power is reduced to 60\% of the original.}}
\label{fig:50}
\end{center}
\end{figure}

\subsection{Trade-off}
\subsubsection{Degradation of SNR and increase in BER}
The advantages of the proposed approach come at the cost of degradation in SNR. As shown in Fig.~\ref{fig:875} \& Fig.~\ref{fig:50}, the required peak in IFFT magnitude decreases to 80\% with \(\beta=0.875\) and to 60\% for \(\beta=0.5\). 
This shows that with the decrease in \(\beta\) the signal component degrades, contributing to higher BER. 
These results are for SF value 7 and SNR of -5 dB.

\begin{figure}[t]
\begin{center}
\centering
\includegraphics[width=0.5\textwidth]{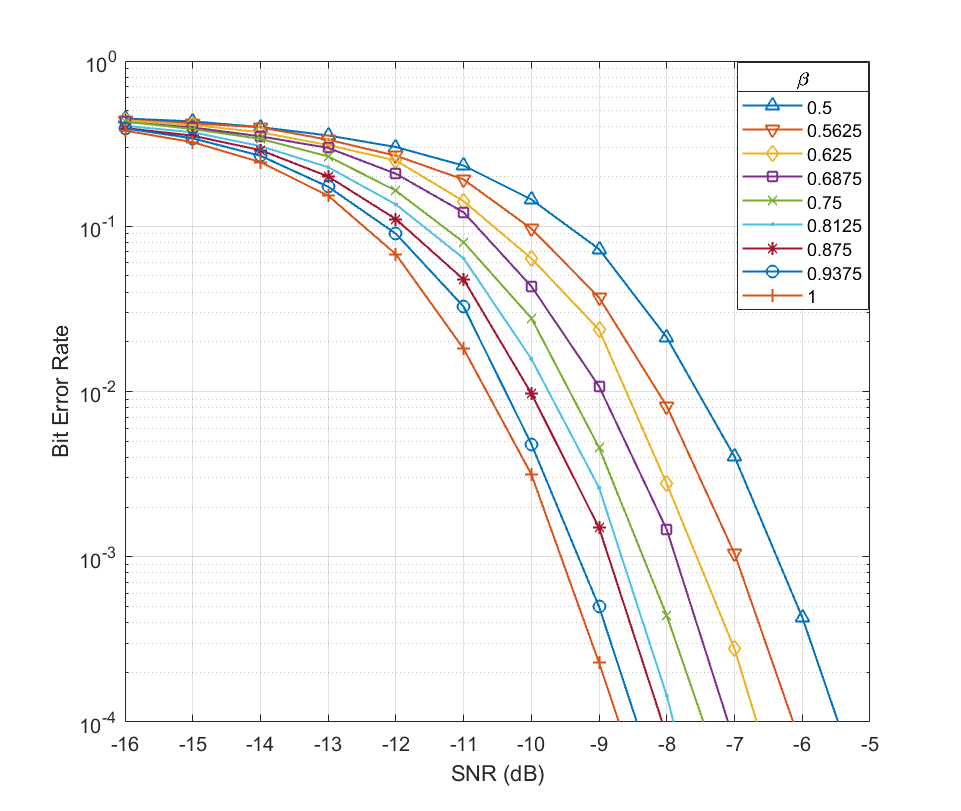}
\caption{{Comparison of bit error rates for different values of reduction factor \(\beta\)  (lower \(\beta\) gives higher BER at same SNR).}}
\label{fig:snr}
\end{center}
\end{figure}

This can also be seen in bit error rate graph in Fig.~\ref{fig:snr}. 
As a decrease in \(\beta\) reduces symbol period and effective symbol energy, this translates to degradation of SNR at the receiver. 
Hence it can be seen that decrease in \(\beta\) increases BER at the receiver.

Thus the margin, by which SNR is higher than the minimum required, should be used to determine the suitable symbol period reduction factor \(\beta\).
This surplus SNR margin required for different values of reduction factor \(\beta\) can be understood from Fig.~\ref{fig:snr}.

Since the proposed approach depends on SNR for deciding symbol period reduction factor \(\beta\), the end devices which do not have SNR higher than the minimum required, should not use the proposed approach. In other words, their \(\beta\) value should be always 1.

\subsubsection{Symbol Overhead} The proposed approach also has an overhead of one extra symbol in header of the packet.
This symbol contains information about the reduction factor \(\beta\) used by the transmitter for subsequent payload symbols.

For the payload of \(N_s\) symbols, the saving in time-on-air is given as \((1-\beta)\times T_s \times N_s\) and the overhead is just one symbol of time period \(T_s\). Thus this overhead is negligible in case of large payloads and small beta value.

% According to LoRa specifications \cite{alliance}, the maximum value of \(N_s\) is defined as 255. So the maximum time saving for the maximum payload becomes \((1-\beta)\times T_s \times 255\).

Therefore, as long as time saving in payload part is greater than 1, it can give time saving for the packet. This time saving for packet \(T_{saving}\) can be quantified as, 

\begin{equation}\label{uk1}
T_{saving} =  N_s \times (1 - \beta) \times T_s  - T_s.
\end{equation}

The subtraction of \(T_s\) in above equation corresponds to the overhead of one symbol added in the header.

\subsubsection{Effect on dense network}
In a dense LoRa deployment, in order to reduce collision and congestion, reducing packet duration becomes more important than optimizing transmission powers of end devices.
% as higher packet duration may result in more collisions and thus poor network throughput.

In the dense network scenario, some of the end devices, which have good SNR conditions, can benefit from the proposed approach and adapt their symbol period to increase their data rate. 
This reduces their packet duration and thus, reduces network congestion contributed by these devices to some extent. 
This results in decrease of collision probability and also improvement in the packet delivery rate.

\subsubsection{Adoption in LoRaWAN}
In order to increase lifetime of the battery operated end devices, LoRaWAN uses ADR mechanism to optimizes transmission powers of end devices. 
%But if the devices are of Class B or Class C this can be neglected. 
But for class C devices which don't need to have any constraint on energy consumption, data rate enhancement holds higher preference than transmission power reduction. 
Thus, the proposed approach can be aggressively employed on such devices to get higher data rate.
This results in the reduction in transmission time of corresponding packets and thus, reduction in the network congestion. 

The proposed approach gives all the improvements mentioned above, but also degrades BER at the receiver by a small margin. 
Thus, similar to the existing ADR mechanism in LoRaWAN, a better way can be to monitor SNR for consecutive packets and employ the proposed approach to adaptively reduce symbol period depending on additional available SNR margin over the minimum required.
% if SNR is better than the minimum required.
% ADR tries to optimize network parameters, but t
Therefore the LoRa end devices having good SNR can use it to their advantage and transmit their data at higher rate using the proposed approach.

%\begin{figure}[htbp]
%\centerline{\includegraphics{fig1.png}}
%\caption{Example of a figure caption.}
%\label{fig}
%\end{figure}

\section{Conclusion}

In this paper, we have proposed an adaptive symbol period based approach for data rate enhancement in LoRa. 
% This is achieved by adaptive reduction of symbol period at physical layer. 
The reduction in symbol period is achieved by truncation of chirp signal corresponding to LoRa payload symbol.
This reduces the time-on-air of packets and therefore, mitigates packet collisions as well as network congestion. 
The proposed approach increases BER at the receiver by a small margin depending on reduction factor \(\beta\). 
However, it is acceptable in case of good communication link conditions having SNR much higher than the minimum required. 
Thus, the proposed approach can be aggressively employed in LoRa devices and the overall network performance can be improved.
% instead of transmission power adaptation through ADR mechanism
% , since data rate enhancement is more beneficial than transmission power reduction for these devices.

\section*{Acknowledgment}
This  work  is  supported  by  Visvesvaraya  Ph.D.  Scheme, Ministry of Electronics and Information Technology (MeitY), Govt. of India and Indian Institute of Technology Hyderabad, India.

\end{document}